\documentclass[prl,superscriptaddress,showpacs,twocolumn]{revtex4}
\usepackage{graphicx}
\usepackage{epsf}

\def\bea{\begin{eqnarray}}
\def\eea{\end{eqnarray}}
\def\beq{\begin{equation}}
\def\eeq{\end{equation}}
\newcommand{\ice}[1]{\relax}

\newcommand\VV{\setbox0=\hbox{V}\hbox{\rm V\raise\ht0
  \hbox to0pt{\hss\vbox to0pt{\hbox{v}\vss}}}}
\def\slashchar#1{\setbox0=\hbox{$#1$}           
   \dimen0=\wd0                                 
   \setbox1=\hbox{/} \dimen1=\wd1               
   \ifdim\dimen0>\dimen1                        
      \rlap{\hbox to \dimen0{\hfil/\hfil}}      
      #1                                        
   \else                                        
      \rlap{\hbox to \dimen1{\hfil$#1$\hfil}}   
      /                                         
   \fi}                                         %

\begin{document}

\title{\mbox{\boldmath $\Lambda_b$} lifetime puzzle in heavy-quark expansion}

\author{Fabrizio Gabbiani}

\affiliation{Department of Physics and Astronomy, 
Wayne State University, Detroit, MI 48201}

\author{Andrei I.~Onishchenko}

\affiliation{Department of Physics and Astronomy, 
Wayne State University, Detroit, MI 48201}

\affiliation{Institute for High Energy Physics, 142284
Protvino, Moscow region, Russia}

\author{Alexey A.~Petrov}

\affiliation{Department of Physics and Astronomy, 
Wayne State University, Detroit, MI 48201}

\affiliation{Michigan Center for Theoretical Physics,
University of Michigan, Ann Arbor, MI 48109}

\begin{abstract}
Lifetime differences of heavy hadrons can be consistently computed 
in heavy-quark expansion. The leading effects appear through spectator 
interactions at order $1/m_b^3$. We compute a well-defined subset of 
$1/m_b^4$ corrections to the lifetime ratio of $\Lambda_b$ baryon 
and $B_d$ meson. We find that these corrections are large and 
should be taken into account in the systematic analysis of heavy hadron
lifetimes. We claim that they could shift the ratio 
$\tau_{\Lambda_b}/\tau_{B_d}$ by as much as $-4.5\%$, significantly reducing 
the discrepancy between the theoretical predictions and experimental 
observations.

\end{abstract}


\maketitle
Barring possible duality violations, one of the most unambiguous
predictions of the heavy-quark effective theory
is a prediction of the ratios of lifetimes of heavy mesons.
In that respect, the low experimental value of the ratio
$\tau(\Lambda_b)/\tau(B_d)$ appears quite puzzling. While the lifetime ratios
of heavy mesons appear to be consistent with the experimentally observed
ones~\cite{PDG},
\bea
\left . \frac{\tau(B_u)}{\tau(B_d)}\right |_{ex}&=&1.074 \pm0.014,
\left . \frac{\tau(B_u)}{\tau(B_d)}\right |_{th}=1.07 \pm 0.03, ~\quad
\nonumber \\
\left . \frac{\tau(B_s)}{\tau(B_d)}\right |_{ex}&=&0.948 \pm 0.038, 
\left . \frac{\tau(B_s)}{\tau(B_d)}\right |_{th}=1.00 \pm 0.02,~\quad
\eea
the latest experimental observations suggest that
\beq
\tau(\Lambda_b)/\tau(B_d)|_{ex}= 0.798 \pm 0.052,
\eeq
which differs rather significantly from the theoretical predictions of
$0.90\pm0.05$~\cite{Bigi:1994wa,Neubert:1996we,Rosner:1996fy,Franco:2002fc}. 
It is therefore worthwhile to look for other effects that affect the baryon 
lifetime while preserving the lifetime ratios of mesons.

Inclusive decay rates can be computed in the heavy-quark expansion.
The most convenient way of doing so is to employ the optical theorem to
relate the decay width to the imaginary part of the matrix element of the 
forward scattering amplitude:
\bea \label{rate}
\Gamma(H_b \to X)&=&\frac{1}{2 M_{H_b}} \langle H_b |{\cal T} | H_b \rangle,
\nonumber \\
{\cal T} &=& {\mbox{Im}}~ i \int d^4 x T \left\{
H_{\mbox{\scriptsize eff}}(x) H_{\mbox{\scriptsize eff}}(0) \right \}.
\eea
Here $H_{\mbox{\scriptsize eff}}$ represents an effective $\Delta B=1$ Hamiltonian
at the scale $\mu = m_b$,
\beq
H_{\mbox{\scriptsize eff}} = \frac{4 G_F}{\sqrt{2}} V_{cb} 
\sum \left[c_1 Q_1^{u'd'} +
c_2 Q_2^{u'd'} \right] + h.c.,
\eeq
where $d'$ and $u'$ are quark flavor eigenstates, $c_i$ are the Wilson coefficients, 
and the four-quark operators $Q_1$ and $Q_2$ are given by
\beq
Q_1^{u'd'}=\bar d_{L}' \gamma_\mu u_{L}' ~\bar c_{L} \gamma^\mu b_{L},~
Q_2^{u'd'}=\bar c_{L} \gamma_\mu u_{L}' ~\bar d_{L}' \gamma^\mu b_{L}.
\eeq
In the heavy-quark limit, the energy release is large and therefore an
Operator Product Expansion (OPE) can be constructed for Eq.~(\ref{rate}),
which results in series of local operators of increasing dimension
suppressed by powers of $1/m_b$. In other words, the calculation of ${\cal T}$
in the expression for the rate in Eq.~(\ref{rate}) is equivalent to computing 
matching coefficients of the effective $\Delta B=0$ Hamiltonian at the scale 
$\mu = m_b$, with subsequent computation of its matrix elements.

At the leading order in the heavy-quark expansion all heavy hadrons have 
the same lifetime. The situation changes at higher orders. At order 
$1/m_b^2$ the difference between meson and baryon lifetimes appear due to the 
difference in their structure. The ratio of lifetimes of $\Lambda_b$ and $B_d$ is
\bea
\frac{\tau(\Lambda_b)}{\tau(B_d)} &=& 1 +
\frac{1}{2 m_b^2} \left [ \mu_\pi^2(\Lambda_b)-\mu_\pi^2(B_d) \right] 
\nonumber \\
&+&
\frac{C_G}{m_b^2} \left [ \mu_G^2(\Lambda_b)-\mu_G^2(B_d) \right] +
{\cal O}(1/m_b^3),
\eea
with $C_G \approx 1.2$~\cite{Bigi:1994wa,Neubert:1996we}. $\mu_\pi^2$ and
$\mu_G^2$ represent kinetic-energy and chromomagnetic interaction 
corrections~\cite{Bigi:1994wa}. At this order in heavy-quark expansion, the
difference is mainly driven by the fact that light quarks in 
$\Lambda_b$ appear in a $J^P=0^+$ quantum state, diminishing any 
correlations of spins between the heavy-quark and light cloud. This implies that 
$\mu_G^2(\Lambda_b)=0$. With matrix elements of kinetic-energy operators canceling 
each other to a large degree, this difference amounts to at most $1-2\%$, which 
is not sufficient to explain the observed pattern of lifetimes. 

The main contribution comes from the dimension-six operators that enter at
the $1/m_b^3$ level. An important
subclass of these operators involves four-quark operators, whose contribution
is additionally enhanced due to the phase-space factor $16 \pi^2$. These
effects are commonly called Weak Annihilation (WA), Weak Scattering (WS), and 
Pauli Interference (PI). They introduce differences in lifetimes of all heavy 
mesons and baryons~\cite{Bigi:1994wa,Neubert:1996we,Shifman:wx,Guberina:1979xw}. 
Their contribution to the lifetime ratios are governed by matrix elements of 
$\Delta B=0$ four-fermion operators
\bea\label{Spec}
{\cal T}_{\rm spec} &=& 
{\cal T}_{\rm spec}^{u} +
{\cal T}_{\rm spec}^{d'} +
{\cal T}_{\rm spec}^{s'},
\eea
where ${\cal T}_i$ contributing to Eq.~(\ref{rate}) are
\bea\label{SpecLO}
{\cal T}_{\rm spec}^{u} &=& 
\frac{G_F^2m_b^2|V_{bc}|^2(1-z)^2}{2\pi}
\biggl\{
\left(c_1^2+c_2^2\right)
O_1^{u}
\nonumber 
\\
&+&2c_1c_2\tilde{O}_1^{u}
+ \delta_{1/m}^{u}
+ \delta_{1/m^2}^{u}
\biggr\}, \\
{\cal T}_{\rm spec}^{d'} &=&
-\frac{G_F^2m_b^2|V_{bc}|^2(1-z)^2}{4\pi}
\biggl\{
c_1^2 \biggl[
(1+z)O_1^{d'}
\nonumber 
\\
&+& \frac{2}{3}(1+2z)O_2^{d'}
\biggr] +\left(
N_cc_2^2+2c_1c_2
\right)
\left[
(1+z)\tilde{O}_1^{d'} \right.
\nonumber \\ 
&& \left.
+\frac{2}{3}(1+2z)\tilde{O}_2^{d'}
\right] 
+\delta_{1/m}^{d'} + \delta_{1/m^2}^{d'}
\biggr\}, \\
{\cal T}_{\rm spec}^{s'} &=&
-\frac{G_F^2m_b^2|V_{bc}|^2\sqrt{1-4z}}{4\pi}
\left\{
c_1^2\left[
O_1^{s'} \right.\right .
\nonumber \\ 
&& \left. \left.
+\frac{2}{3}(1+2z)O_2^{s'}
\right] 
+\left(N_cc_2^2+2c_1c_2
\right)
\left[
\tilde{O}_1^{s'} \right. \right.
\nonumber \\ 
&& \left. \left.
+\frac{2}{3}(1+2z)\tilde{O}_2^{s'}
\right]
+\delta_{1/m}^{s'} + \delta_{1/m^2}^{s'}
\right\}.
\eea
$O_i^q$ are the four-fermion operators
\bea\label{LeadingOps}
O_1^q &=& \bar b_i\gamma^{\mu}(1-\gamma_5)b_i\bar q_j\gamma_{\mu}(1-\gamma_5)q_j, 
\nonumber \\
O_2^q &=& \bar b_i\gamma^{\mu}\gamma_5b_i\bar q_j\gamma_{\mu}(1-\gamma_5)q_j,
\eea
with $z=m_c^2/m_b^2$. The $\tilde{O}_i^q$ denote the color-rearranged
operators that follow from the expressions for $O_i^q$ by
interchanging the color indexes of the $b_i$ and $q_j$ Dirac spinors.
Our choice of basis operators differs from the one made 
in~\cite{Neubert:1996we} due to the relative simplicity of the higher 
order operators in our basis. $\delta_{1/m}^{q'}$ and $\delta_{1/m^2}^{q'}$
represent power-suppressed corrections to the spectator 
contributions. These corrections are clearly parametrically more important 
than the charm-quark mass effects normally taken into account~\cite{Neubert:1996we}, 
as the latter are suppressed by $z$ or {\it two powers} of $1/m_b$. This subset 
of the full set of $1/m_b^4$ corrections also retains the $16\pi^2$ ``phase-space'' 
enhancement enjoyed by the leading effects of WS and PI. However,
they need not to interfere destructively in their contribution to the $\Lambda_b$ 
lifetime as do WS and PI~\cite{Rosner:1996fy}. Conspiring, they can produce 
a sizable shift in the ratio of the $\Lambda_b$ and $B$-meson lifetimes. We shall argue 
below that this is indeed the case.

Most of the recent progress in understanding lifetimes has been concentrated on 
computing the next-to-leading order (NLO) QCD corrections to Wilson
coefficients of the operators in Eq.~(\ref{SpecLO})~\cite{Beneke:2002rj,Franco:2002fc} 
or calculating matrix elements of these operators in quark models or on the lattice. 
The simplest parameterization of these matrix elements is inspired by a
naive factorization ansatz and represents the matrix elements of the four-fermion operators 
as products of two matrix elements of current operators separated by a vacuum state. 
This procedure introduces four new scale-dependent parameters
$B_i(\mu)$ and $\epsilon_i(\mu)$~\cite{Neubert:1996we} which in our basis parametrize
the matrix elements of the operators in Eq.~(\ref{LeadingOps}) as
\bea
\langle B_q |O_1^q| B_q \rangle &=& f^2_{B_q} m^2_{B_q} \left(2 \epsilon_1 + \frac{B_1}{N_c}
\right), \nonumber \\
\langle B_q |\widetilde{O}_1^q| B_q \rangle &=& f^2_{B_q} m^2_{B_q} B_1,
\\
\langle B_q |O_2^q| B_q \rangle &=& -f^2_{B_q} m^2_{B_q} \left[
\frac{m_{B_q}^2}{m_b^2} \left(2 \epsilon_2 + \frac{B_2}{N_c}\right)
\right.\nonumber \\
&+& \left.
\frac{1}{2}\left(2 \epsilon_1 + \frac{B_1}{N_c}
\right)
\right],
\nonumber \\
\langle B_q |\widetilde{O}_2^q| B_q \rangle &=& -f^2_{B_q} m^2_{B_q} \left[
\frac{m_{B_q}^2}{m_b^2} B_2 + \frac{1}{2} B_1
\right].
\nonumber
\eea
Similar expressions are available for baryons where they are motivated 
by the valence quark model
\begin{figure}[t]
\centerline{\epsfxsize=6.5cm\epsffile{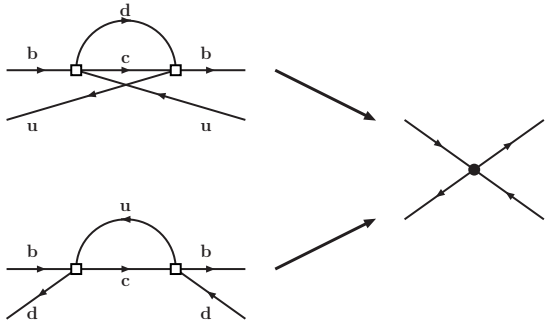}}
\centerline{\parbox{7.5cm}{\caption{\label{fig:leading}
Higher order $1/m_b$  and $1/m_b^2$ corrections to 
spectator contributions (derivative insertions).}}}
\end{figure}
\bea
\langle \Lambda_b |O_1^q| \Lambda_b \rangle &=& 
-\widetilde{B} \langle \Lambda_b |\widetilde{O}_1^q| \Lambda_b \rangle = 
\frac{\widetilde{B}}{6}  f^2_{B_q} m_{B_q} m_{\Lambda_b} r, \quad
\nonumber \\
\langle \Lambda_b |O_2^q| \Lambda_b \rangle &=& 
-\widetilde{B} \langle \Lambda_b |\widetilde{O}_2^q| \Lambda_b \rangle=
\frac{\widetilde{B}}{6}
f^2_{B_q} m_{B_q} m_{\Lambda_b} \delta, \quad
\eea
where $r=\left|\psi^{\Lambda_b}_{bq}(0)\right|^2/\left|\psi^{B_q}_{b\bar q}(0)\right|^2$
is the ratio of the wave functions at the origin of the $\Lambda_b$ and $B_q$ mesons, and
$\widetilde{B}=1$ in the valence-quark model. Estimates of $r$ vary from 
0.1 to 1.8 and can potentially be larger~\cite{Neubert:1996we}.
Note that $\delta={\cal O}(1/m_b)$, which follows from the heavy-quark spin symmetry.
The above parameters can be computed in QCD sum rules, quark models, or on the lattice. 
Naively, one expects that in the large-$N_c$ limit $B_1\sim B_2 ={\cal O}(1), 
~\epsilon_1\sim \epsilon_2 = {\cal O}(1/N_c)$. Yet, the contributions
of the ``octet'' parameters $\epsilon_i$ are important due to the large 
Wilson coefficient that accompanies them and the (accidental) cancellation that
suppresses the Wilson coefficient accompanying the $B_i$ parameters. A compilation of 
various estimates of these parameters can be found in~\cite{Chay:1999pa}. 
One can parametrize the meson-baryon lifetime ratio as
\bea \label{MesonBaryon}
\frac{\tau(\Lambda_b)}{\tau(B_d)} \simeq  0.98 - (d_1+d_2 \widetilde{B})r
-(d_3 \epsilon_1+d_4 \epsilon_2) \nonumber \\
-(d_5 B_1+d_6 B_2) 
+ \delta_{1/m},\qquad
\eea
where the scale-dependent parameters ($d_i(m_b)=\{0.012, 0.021, 0.173, -0.195,
{\cal O}(10^{-3}),{\cal O}(10^{-3})\}$ at LO~\cite{Neubert:1996we} and 
$\{0.023, 0.028, 0.16, -0.16, 0.08, -0.08\}$ at NLO~\cite{Franco:2002fc})
are defined in~\cite{Neubert:1996we}. While tempting, it is clearly difficult to 
reduce the $\tau(\Lambda_b)/\tau(B_d)$ lifetime ratio by inflating $\epsilon_i$ 
without disturbing the meson lifetime ratios.
Thus, at least at the $1/m_b^3$ level, the problem can be ameliorated by conjecturing
that $r\gg 1$, which runs in contrast with otherwise successful quark-model 
expectations. $\delta_{1/m}$ represents contributions of order $1/m_b^4$ and higher, 
which we shall address. The impact of $1/m_b^4$ corrections can be naively expected 
at the level of $20\%$. However, as we shall see below, kinetic corrections to 
WS and PI {\it conspire} in $\Lambda_b$ and, coupled with large Wilson 
coefficients, produce a sizable effect. We computed the higher order corrections 
to Eq.~(\ref{Spec}) in the heavy-quark expansion, denoted below as
$\delta_{1/m}^{q}$ and $\delta_{1/m^2}^{q}$, by expanding the forward scattering 
amplitude in Eq.~(\ref{rate}) in the light-quark 
momentum and matching the result onto the operators containing derivative insertions 
(see Fig.~\ref{fig:leading}). The result is
\begin{figure}[t]
\centerline{\epsfxsize=6.3cm\epsffile{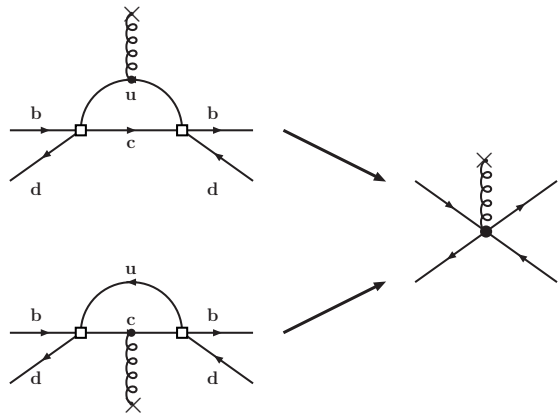}}
\centerline{\parbox{7.5cm}{\caption{\label{fig:subleading}
Higher order $1/m_b^2$ corrections to spectator contributions 
(background gluon interactions).}}}
\end{figure}
\bea \label{OurCorrection}
\delta_{1/m}^{u} &=&
-2\left(c_1^2+c_2^2\right)R_1^{u}
-4c_1c_2 \tilde{R}_1^{u},  \nonumber \\
\delta_{1/m}^{d',s'} &=&
\frac{2}{3} c_1^2 \left[
R_1^{d',s'}+
R_2^{d',s'}-R_3^{d',s'}
\right] \\ 
&+&\frac{2}{3}\left(N_cc_2^2+2c_1c_2\right)
\left[
\tilde{R}_1^{d',s'}+
\tilde{R}_2^{d',s'}-\tilde{R}_3^{d',s'}
\right].
\nonumber 
\eea
The operators $R_i$ are defined as
\bea
R_1^q &=& \frac{1}{m_b^2}\bar b_i\gamma^{\mu}(1-\gamma_5)\vec{D}^{\alpha}b_i
\bar q_j\gamma_{\mu}(1-\gamma_5)\vec{D}_{\alpha}q_j, \nonumber \\
R_2^q &=& \frac{1}{m_b^2}\bar b_i\gamma^{\mu}(1-\gamma_5)\vec{D}^{\nu}b_i
\bar q_j\gamma_{\nu}(1-\gamma_5)\vec{D}_{\mu}q_j \nonumber \\
R_3^q &=& \frac{m_q}{m_b}\bar b_i(1-\gamma_5)b_i\bar q_j(1-\gamma_5)q_j. 
\eea
Here $\tilde{R}_i^q$ denote the color-rearranged operators that follow from 
the expressions for $R_i^q$ by interchanging the color indexes of $b_i$ and $q_j$ Dirac 
spinors. We dropped all the contributions suppressed by light and charm-quark masses, except for
$R_3^q$. Since the above result contains ``full'' QCD $b$-fields, no immediate power counting 
for these operators is available. The power counting becomes manifest at the level of 
the matrix elements. We shall present the most general parametrization of these 
matrix elements elsewhere. 
Neglecting non-factorizable contributions, the meson matrix elements are
\bea \label{MEMeson}
\langle B_q |R_1^q| B_q \rangle &=& 
\langle B_q |\widetilde{R}_1^q| B_q \rangle/N_c
\nonumber \\
 &=& \frac{\beta_1}{2 N_c} f^2_{B_q} m^2_{B_q}
\left[\frac{m_{B_q}^2}{m_b^2}-1\right], \nonumber \\
\langle B_q |R_{2,3}^q| B_q \rangle &=& 
-\frac{\beta_{2,3}}{4N_c} f^2_{B_q}
\left(m_{B_q}^2-m_b^2\right),
\nonumber \\
\langle B_q |\widetilde{R}_{2,3}^q| B_q \rangle &=& 
-\frac{\beta_{2,3}}{4} f^2_{B_q}
\left(m_{B_q}^2-m_b^2\right),
\eea
where parameters $\beta_i=1$ in the factorization approximation~\cite{Beneke:1996gn}, 
which we shall employ hereafter. Similarly, we used 
the quark-diquark model to {\it guide} our parametrizations of baryon matrix 
elements
\bea \label{MEBaryon}
\langle \Lambda_b |R_1^q| \Lambda_b \rangle &=& 
- \langle \Lambda_b |\widetilde{R}_1^q| \Lambda_b \rangle \nonumber \\
&=& - \frac{\tilde{\beta_1}}{24}
f^2_{B_q} m_{B_q} m_{\Lambda_b} 
\left[\frac{m_{\Lambda_b}^2}{m_b^2}-1\right], \nonumber \\
\langle \Lambda_b |R_2^q| \Lambda_b \rangle &=& 
- \langle \Lambda_b |\widetilde{R}_2^q| \Lambda_b \rangle \\
&=& -\frac{\tilde{\beta_2}}{48 m_b^2}
f^2_{B_q} \frac{m_{B_q}}{m_{\Lambda_b}}
\left(m_{\Lambda_b}^4-m_b^4\right), \nonumber \\
\langle \Lambda_b |R_3^q| \Lambda_b \rangle &=& 
-\langle \Lambda_b |\widetilde{R}_3^q| \Lambda_b \rangle =
-\frac{\tilde{\beta_3}}{\tilde{\beta_2}}\langle \Lambda_b |R_2^q| \Lambda_b \rangle,
\nonumber
\eea
where $\tilde{\beta_i}=r$ in the approximation where the color of the quark 
fields in the operators matches the color of the quarks inside the baryon, which is
an analogue of the factorization approximation for baryons.
Inserting Eqs.~(\ref{MEMeson}) and (\ref{MEBaryon}) into (\ref{OurCorrection}) 
gives an estimate of our correction. Numerically, it constitutes $40-60\%$ of the leading 
spectator contribution if the leading logarithmic approximation is employed for WS and PI,
depending on the chosen renormalization scale (we varied the scale from $m_b/2$ to $2 m_b$). 
Employing the full NLO result for WS and PI~\cite{Franco:2002fc} we observe that the effect 
of $1/m^4$ corrections reduces to $36-45\%$ of the leading spectator contribution.
While such a sizable effect is surprising, the main  
source of such a large correction can be readily identified, at least in 
factorization. While the individual $1/m_b$ corrections to WS and PI are of order
$20\%$, as expected from the naive power counting, they contribute to
the $\Lambda_b$ lifetime with the same sign, instead of destructively interfering WS and 
PI~\cite{Rosner:1996fy}. This conspiracy of several small $\sim 20\%$ 
effects produces a sizable shift in the ratio of the $\Lambda_b$ and $B$-meson lifetimes,
which can be as large as $-4.5\%$. We expect this effect to persist with more rigorous 
computations of matrix elements as well. 

We checked that higher order $1/m_b^5$ contributions are under control and 
as large as one would expect based on naive power counting, i.e. of the
order of a few percent. These higher order contributions arise from graphs
with more derivative insertions and interactions with background gluon fields.
Discarding light and charm-quark masses we obtain: 
\bea
\delta_{1/m^2}^{u} = 0, ~~  
\delta_{1/m^2}^{d'(s')} &=& \frac{c_1^2}{m_b^2} \left[
P_1^{d'(s')}-P_2^{d'(s')}
\right],
\eea
where the only non-zero contribution comes from the gluonic operators depicted in
Fig.~\ref{fig:subleading}
\bea \label{OneOverMbSquared}
P_1^q &=& \bar b_i \gamma^\mu (1-\gamma_5) \widetilde{G}_{\mu\nu} b_k
\bar d_j \gamma^\nu (1-\gamma_5) d_j, \nonumber \\
P_2^q &=& \bar b_i \gamma^\mu (1-\gamma_5) b_i
\bar d_j \gamma^\nu \widetilde{G}_{\mu\nu} (1-\gamma_5) d_k,
\eea
where $\widetilde{G}_{\mu\nu} = t_{ik}^a \widetilde{G}^a_{\mu\nu}$. It is easy to see 
that the naive power counting for the matrix elements of the operators in 
Eq.~(\ref{OneOverMbSquared}) implies that $1/m_b^2$ contribution to the
spectator effects is of the order of a few percent.

In conclusion, we computed a well-defined subset of $1/m_b^4$ corrections to the lifetime 
ratio of $\Lambda_b$ and $B_d$. While this subset does not dominate the full 
$1/m_b^4$ correction in any limit, it receives the same phase-space enhancement factor 
as the leading spectator effect. We found this correction to be large, of order
$40-60\%$ of the leading spectator effect at LO~\cite{Bigi:1994wa,Neubert:1996we,Rosner:1996fy}
and $36-45\%$ at NLO~\cite{Franco:2002fc}, reducing the lifetime ratio by as much
as $4.5\%$ in addition to the ${\cal O}(10\%)$ effect reported earlier. 
This significantly reduces the discrepancy between the theoretical predictions and 
experimental observations, making them compatible within error bars.

{\em Acknowledgments:}
We would like to thank Gene Golowich for helpful discussions. This work was 
supported in part by the US Department of Energy under grant DE-FG02-96ER41005.

\end{document}